\documentclass[a4paper,fleqn,usenatbib]{mnras}

\usepackage[T1]{fontenc}
\usepackage{ae,aecompl}

\usepackage{subfig}
\usepackage{xcolor}

\usepackage{graphicx}	% Including figure files 
\usepackage{amsmath}    % Advanced maths commands
\usepackage{amssymb}    % Extra maths symbols

\title[Gas-phase metallicity gradients and star formation efficiency]{The evolution of the metallicity gradient and the star formation efficiency in disc galaxies}

\author[Sillero et al.]{Emanuel Sillero$^{1}$\thanks{E-mail: esillero@oac.unc.edu.ar}, 
        Patricia B. Tissera$^{2}$, 
        Diego G. Lambas$^{1}$, 
        Leo Michel-Dansac$^{3}$\\
$^{1}$ Instituto de Astronom\'ia Te\'orica y Experimental (CONICET-UNC), Laprida 925, C\'ordoba, Argentina.\\
$^{2}$ Departamento de Ciencias F\'isicas, Universidad Andres Bello, Av. Rep\'ublica 220, Santiago, Chile. \\
$^{3}$ Univ Lyon, ENS de Lyon, Univ Lyon1, CNRS, Centre de Recherche Astrophysique de Lyon UMR5574, F-69007, Lyon, France.}

\date{Accepted XXX. Received YYY; in original form ZZZ}

\pubyear{2017} % Enter the current year, for the copyright statements etc.

\begin{document}

\label{firstpage}
\pagerange{\pageref{firstpage}--\pageref{lastpage}}

\maketitle

\begin{abstract}
    We study the oxygen abundance profiles of the gas-phase components in hydrodynamical simulations of pre-prepared disc galaxies including major mergers, close encounters and isolated configurations.
    We analyse the evolution of the slope of oxygen abundance profiles and the specific star formation rate (sSFR) along their evolution.
    We find that galaxy-galaxy interactions could generate either positive and negative gas-phase oxygen profiles depending on the state of evolution.
    Along the interaction, galaxies are found to have metallicity gradients and sSFR consistent with observations, on average.
    Strong gas inflows produced during galaxy-galaxy interactions or as a result of strong local instabilities in gas-rich discs are able to produce both a quick dilution of the central gas-phase metallicity and a sudden increase of the sSFR.
    Our simulations show that, during these events, a correlation between the metallicity gradients and the sSFR can be set up if strong gas inflows are triggered in the central regions in short timescales.
    Simulated galaxies without experiencing strong disturbances evolve smoothly without modifying the metallicity gradients.
    Gas-rich systems show large dispersion along the correlation.
    The dispersion in the observed relation could be interpreted as produced by the combination of galaxies with different gas-richness and/or experiencing different types of interactions.
    Hence, our findings suggest that the observed relation might be the smoking gun of galaxies forming in a hierarchical clustering scenario.
\end{abstract}


\section{Introduction}

    Observational results show that the cosmic star formation density increases from high redshift to around $z \sim 2$ from where it starts to decline \citep{madau2014}.
    Several mechanisms contribute to the regulation of the star formation activity in galaxies as a function of time.
    Supernova and AGN feedbacks can regulate the star formation activity in galaxies of different masses \citep[][]{scan08,gibson2013,angles-alcazar2014,rosas-guevara2015,crain2015,tissera2016a,muratov2017}.
    These mechanisms act on local and global scales modifying the physical properties of the gas in condition to form stars, mixing gas clouds of different metallicities and exchanging material with the circumgalactic medium.
    Their joint action can affect the evolution of galaxies properties such as stellar mass, colours, metallicity, morphologies and establish correlations between them.
    Mergers and interactions can also contribute by modifying the gas properties, mixing chemical elements and triggering strong star formation activity \citep{sersic1968, tinsley1978, barton2000, lambas2003, woods2006, ellison2010, patton2011}.

    The chemodynamical properties of the stellar populations and the interstellar medium store important information on the action of these processes along galaxy assembly.
    It is well-known that the gas-phase abundance of galaxies correlates with luminosity \citep[e.g.][]{lequeux1979, dutil1999} and stellar mass \citep[e.g.][]{tremonti2004}.
    In the Local Universe, negative metallicity profiles are measured in disc galaxies, on average \citep[e.g.][]{zaritsky1994,van-zee1998}.
    \citet{ho2015} report a correlation with stellar mass so that low stellar-mass galaxies have metallicity profiles with steeper negative slopes than massive galaxies. 
    Recent results from CALIFA and MaNGA surveys show a more complex dependence on stellar mass \citep{sanchez2014,perez-montero2016, belfiore2017} that, in part, could be ascribed to the larger galaxy samples.
    At high redshift, observations show the existence of more positive gradients for stellar mass galaxies \citep{stott2014,wuyts2016}. 
    A correlation between metallicity gradients and stellar mass can be understood within the context of an inside-out scenario for disc formation where star formation rate and metal production are mainly linked to the gas density on the discs.
    Theoretical and numerical models for galaxy formation support this picture \citep{chiap1997, pilkington2012, calura2012, tissera2014}.
    However, there are other mechanisms which can alter or modify the gas-phase metallicity abundances such as mergers and interactions \citep[e.g.][]{dansac2008} and SN feedback \citep{gibson2013}.
    When comparing results of simulations run with different codes, it is important to bare in mind that they resort to different subgrid physics \citep{scan12}.
 
    Observations of interacting galaxies show to have almost flat or even positive metallicity gradients of the discs and low central abundances at a given stellar mass compared to the mass-metallicity relation \citep{kewley2006, dansac2008, kewley2010}.
    These trends have been reproduced by numerical simulations which showed the action of tidal fields capable of driving internal instabilities that generate central low-metallicity gas inflows \citep{rupke2010a, perez2011,torrey2012}.
    The formation of clumps in gas-rich discs \citep{bournaud2011,perez2011,tacchella2016} as well as low-metallicity gas inflows along filaments \citep{ceverino2016} could also contribute to modify the metallicity gradients.

    Recently, \citet{stott2014} combine the gas-phase metallicity gradient (S(O/H)) with the sSFR of star-forming galaxies, claiming the possible existence of a correlation between them, so that positive gradients would be associated to higher star forming systems (high sSFR galaxies) based on the results reported for interacting galaxies.
    These authors speculate that mergers, interactions or efficient gas accretion into the central regions might be responsible of this correlation as these processes could trigger low-metallicity gas inflows.
    For this analysis, they combined observations of metallicity gradients estimated in wide range of redshifts.
    From a numerical point of view, \citet[][hereafter, T2016]{tissera2016a} observed that galaxies formed in hierarchical clustering universe have gas-phase metallicity slopes and sSFR in good agreement with the observational relation reported by Stott et al.
    These authors also studied simulated galaxies with positive and very negative metallicity slopes finding that both could be ascribed to effects of mergers and close interactions which were able to trigger low-metallicity gas inflows.
    \citet{ma2017} reported a similar correlation using a different numerical code and SN feedback model.
    Previous numerical studies have showed how mergers and interactions can trigger gas inflows if there is enough gas in the systems \citep[e.g.][]{bh96, mh96, tissera2000} and how these gas inflows can affect the metallicity gradients diluting the central abundances \citep{perez2011, rupke2010} {that might} be enriched again by the new-born stars \citep{perez2011}.

    The reported S(O/H)-sSFR correlation suggests a close relationship between the mechanisms that affect both the star formation activity and the metallicity gradient at the same time.
    T2016 showed the variation of the metallicity gradients and the sSFR along a gas-rich major merger which was used as illustration of the impact of this kind of violent event in shaping a correlation between these parameters.
    Considering that in a hierarchical clustering scenario, mergers and interactions are common events in the life of galaxies of different masses, it is then relevant to analyse the evolution of the S(O/H) and the sSFR in detail for different interacting configurations with the purpose of investigate under which conditions the S(O/H)-sSFR relation might be established.
    Considering the variation in subgrid physics used in numerical codes, this relation could provide another way to constrains them.

    In this paper, we studied the effects of mergers on the gas distribution, star formation activities and gas-phase metallicity gradients using at set of idealized simulations so that the impact parameters and the initial conditions of the mergers as well as the characteristics of the galaxies, such as gas fraction, masses, are fully controlled.
    We analysed interactions with different orbital parameters including orbital configurations which produced mergers and fly-by events. Gas-rich and gas-poor configurations are considered.
    The isolated counterparts of the main galaxies have been also run and analysed for comparison.
    We did not aim at covering all possible parameter space but to analyse key ones which allow us to envisage a possible origin for a S(O/H)-sSFR relationship.

    This paper is organised as follows. 
    In Section 2 we describe the main characteristics of the simulations and the galaxy sample.
    In Section 3 we discuss the metallicity gradients and sSFR for simulated disc galaxies and confront them with observations.
    Our main findings are summarised in the conclusions.

\section{Numerical simulations and simulated galaxies}

    We use a set of simulated mergers, fly-by events and isolated galaxies performed by using a version of the code {\small P-GADGET-3}, which includes treatments for metal-dependent radiative cooling, stochastic star formation (SF), chemical enrichment, and the multiphase model for the ISM and the SN feedback scheme of \citet{scan05, scan06}.
    This SN feedback model is able to successfully trigger galactic mass-loaded winds without introducing mass-scale parameters or kicking gas particles, or suppressing hydrodynamical interactions with surrounding particles, for example.
    As a consequence, galactic winds naturally adapt to the potential well of the galaxy where star formation takes place.

    The SN feedback model works within a multiphase model for the ISM which allows the coexistence and material exchange between the hot, diffuse phase and the cold, dense gas phase \citep{scan06, scan08}.
    Stars form in dense and cold gas clouds and part of them ends their lives as SNe, injecting energy and chemical elements into the ISM. 
    The thermodynamical and chemical changes are introduced on particle-by-particle basis and considering the physical characteristics of its surrounding gas medium.
    The detail description and tests of the multiphase medium and the SN feedback are presented in \citet{scan06}.
    Briefly, it is assumed that each SN event releases energy, which is distributed between the cold, dense and hot, diffuse phases.
    We assume that 50 percent of the energy is injected into the cold phase surrounding the stellar progenitor (the rest is injected into the surrounding hot phase).
    \citet{scan08} reported this value to provide the best description of the energy exchange with the ISM for their scheme.
    This energy is thermalized instantaneously in the hot phase. 
    Conversely, the cold phase stores the injected energy in a reservoir until the accumulated is enough to change the entropy of the cold gas particle so that it joins its hot phase.
    Hence, it is a self-regulated process and does not involved sudden changes in kinetic energy.
    The mass-loaded galactic winds induced by the SN model of \citet{scan08} self-regulates according to the potential wells of the galaxies.

    We use the chemical evolution model developed by \citet{mosconi2001}.
    This model considers the enrichment by SNeII and SNeIa adopting the yield prescriptions of \citet{WW95} and \citet{iwamoto1999}, respectively.
    A set of 13 isotopes are followed in time such as O$^{16}$.
    For this study we estimate the number abundance ratios 12 + log(O/H) as a function of time.
    We will take this abundance ratio to track the evolution of the metallicity profiles.
    The distributions of chemical elements in the gas-phase medium and the stellar populations of galaxies reproduce global observational results as shown by \citet{tissera2016a, tissera2016b}.

\subsection{The simulated galaxies}
    To explore the evolution of the gradient (or slopes) of the metallicity profiles and the sSFR during a galaxy-galaxy interactions, we analysed a gas-rich major merger (50 per cent gas in the disc component) and three of the gas-poor mergers (20 per cent gas in the disc component) examined by \citet{perez2011}.
    All of them involved two equal-mass galaxies (1:1).
    In addition we analysed a fly-by event involving two similar mass galaxies (1:1) and run isolated versions of the galaxies used in the gas-rich and gas-poor major mergers.
    Table \ref{table1} provides the principal characteristics of these events.

    \begin{table}
        \centering
        \caption{Initial conditions for the simulated galaxies. Type of event (merger, fly-by, isolated), relative rotation, gas fraction, energy per SN}
        \begin{tabular}{lcccc}    %four columns, alignment for each
            \hline
            Simulation & i$(^{\circ})$  & f$_{\rm gas}$ & E$_{\rm SN}$ & Type \\
            \hline
            SIII       & 0              & 0.20          & 0.5          & Merger \\
            SV         & 0              & 0.20          & 0.1          & Merger \\
            SVI        & 180            & 0.20          & 0.5          & Merger \\
            SIII-f50   & 0              & 0.50          & 0.5          & Merger \\
            \hline
            SI         & 0              & 0.20          & 0.5          & Fly-bye \\
            \hline
            SI-I       & 0              & 0.20          & 0.5          & Isolated \\
            SIII-f50-I & 0              & 0.50          & 0.5          & Isolated \\
            \hline
        \end{tabular}
        \label{table1}
    \end{table}

    Each galaxy in the simulated mergers events is formed by a dark matter halo of $1.29 \times 10^{12}$ M$_{\odot}$, a bulge and disc components of $1.86 \times 10^{10}$ M$_{\odot}$ and $5.58 \times 10^{10}$ M$_{\odot}$ (gas and stars), respectively.
    The gas-poor galaxy is resolved with 100000 dark matter particles of $ 6.43 \times 10^6$ M$_{\odot}$, 45000 star particles of $1.5 \times 10^6$ M$_{\odot}$, in the disc and bulge components and, initially, 50000 gas particles of $3 \times 10^5$ M$_{\odot}$ in the disc components. 
    The gas-rich galaxies are resolved with 200000, 100000 and 100000 initial particles, respectively.
    The gravitational softenings adopted are $ 0.16$ kpc for the gas particles, $0.32$ kpc for dark matter and $0.20$ kpc for the star particles.
    The initial dark matter profiles are consistent with a NFW profile \citep{nfw96} with a virial circular velocity of 160 kms$^{-1}$.
    The discs components follow an exponential profile with a scale-length of $ 3.46$ kpc while the bulges are consistent with a Hernquist profile.
    The analysed mergers have elliptical orbits with a circularity parameter of $\epsilon \sim 0.20$ and a pericentre distance of $\sim 20 $ kpc. 
    This configuration is consistent with those obtained from cosmological simulations \citep{khochfar2006}.

    We analysed a co-rotating (SIII) and a counter-rotating (SVI) gas-poor configurations.
    And we also studied a version of the co-rotating merger with a lower energy feedback to assess at what extent this mechanism is affecting the results (SV).
    The name of the galaxies are set following \citet{perez2011}.
    These authors analysed in detail the evolution of the metallicity gradients during the mergers but did not studied the relation between the variation of the slopes and the SFR.
    The same galaxy was used to build up the fly-by (SI) event.

    The gas-rich merger (SIII-f50) has the same configuration that the SIII run, being the gas fraction the only difference.
    The galaxies of the gas-poor (SIII) and gas-rich (SIII-f50) mergers were run in isolation for comparison.
    The main parameters are summarised in Table \ref{table1}.

    An important aspect to take into account is that the initial metallicity profiles of the disc galaxies have been set ad hoc by adopting a typical value of disc galaxies considering observational results.
    In the case of the gas-poor galaxy, the initial conditions has a metallicity slope of $\sim -0.08$ dex kpc$^{-1}$\citep{dutil1999} with a central oxygen abundance of 12 + log(O/H) = 9.2 \citep{tremonti2004}. 
    The gas-rich galaxy has a steeper initial metallicity gradient of $ -0.1$dex kpc$^{-1}$ and the central abundance is determined so that they are consistent with the mass-metallicity relation at $z \sim 2$ \citep[12 + log O/H = 8.5, ][]{maiolino2008}.
    Hence, these initial oxygen slopes should be taken only as indicative.
    The relevant aspect to consider here is the change of the slope of the abundance profiles as a function of the sSFR, along the interactions.

\section{Metallicity gradients and star formation of the disc galaxies}

    In order to compute the 12 + log(O/H) abundance profiles, concentric spherical shells of 1 kpc radius are adopted.
    We consider gas particles dominated by rotation, disregarding those that are part of galactic outflows with large velocities perpendicular to the disc plane.
    We assume a inner cut-off radius of 1 kpc which is approximately three times the maximum physical gravitational softening. 
    For the maximum radius we adopted two different values: 9 kpc and 12 kpc.
    The results do not change significantly but the dispersion of the relations increases when the outskirts of discs are included, particularly during the close interactions of galaxy pairs.
    Hence, we focus on the results obtained by using a maximum radius of 9 kpc while in Section 3.4 we discuss the impact of incorporating the outer regions.
    Linear regressions were fitted to these metallicity profiles to obtain both the zero point and slope of metallicity gradient.
    We also calculate the star formation rate, the stellar mass and gas inflows as a function of time.
    For the latter, we search for the gas particles that move inside each radial shell as a function of time. 

    This procedure is applied to each of the simulations summarised in Table \ref{table1}, in order to explore the existence of a relation between the metallicity gradients and the star formation activity.

    \subsection{Major mergers}

\begin{figure*}
            \resizebox{17.cm}{!}{\includegraphics{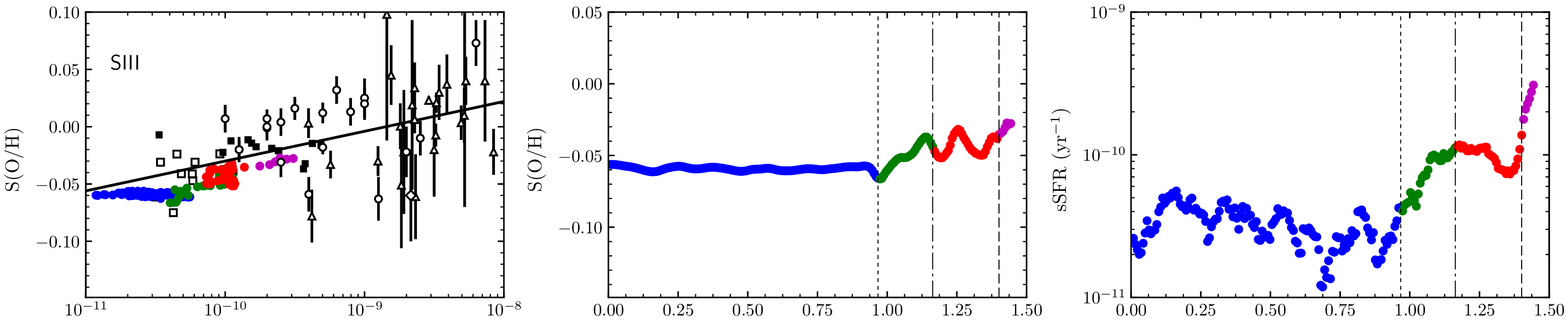}}\\
            \vspace*{-0.2cm}
            \resizebox{17.cm}{!}{\includegraphics{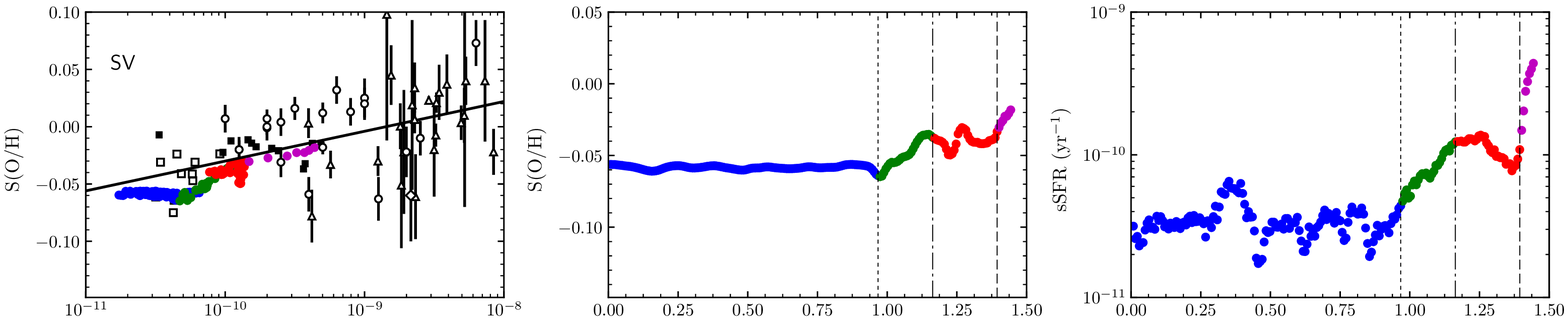}}\\
            \vspace*{-0.2cm}
            \resizebox{17.cm}{!}{\includegraphics{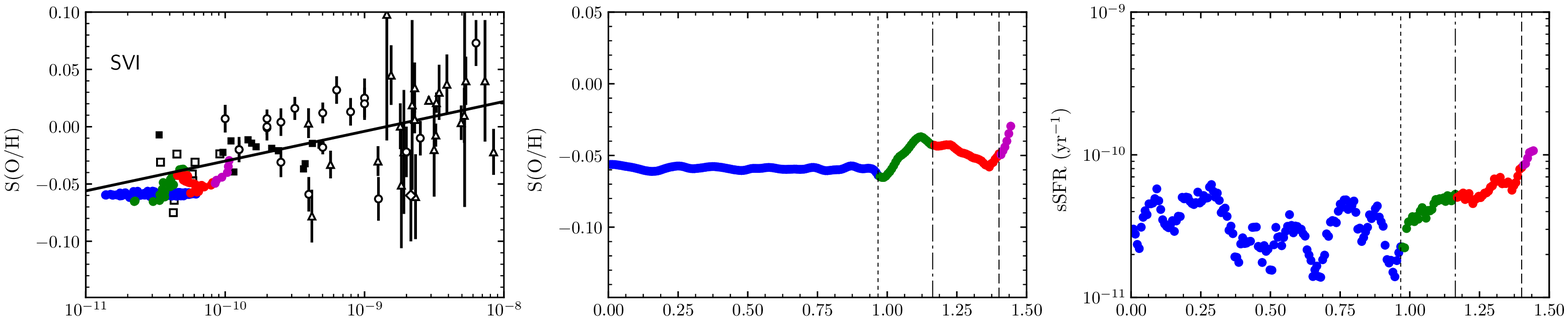}}\\
            \vspace*{-0.2cm}
            \resizebox{17.cm}{!}{\includegraphics{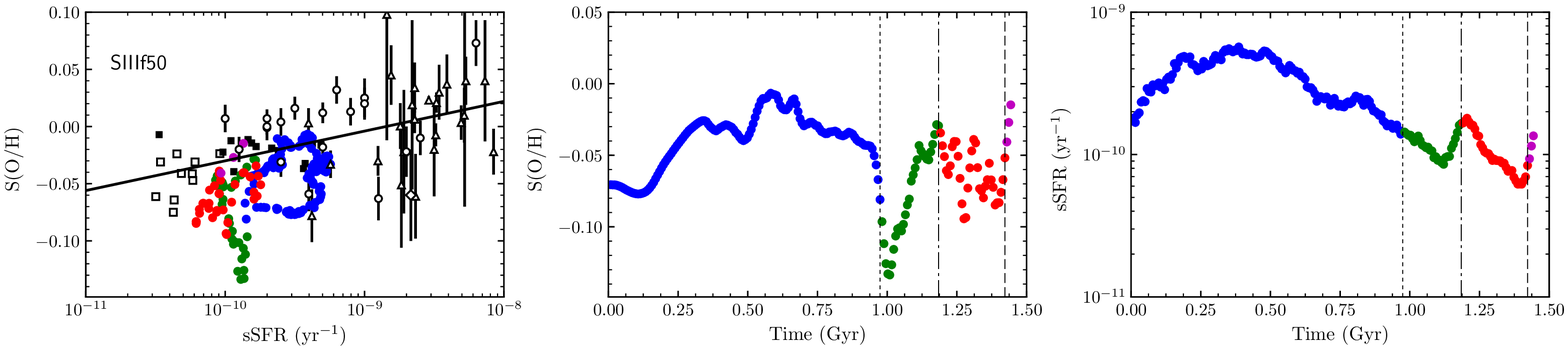}}\\
            \caption{Mean slope of the gas-phase oxygen abundance profiles as a function of the sSFR (left panels) and both quantities as a function of time (middle panels: gas phase oxygen slope; right panels: sSFR) for the simulated galaxies in the gas-poor simulations: SIII (co-rotating configuration), SVI (counter-rotation configuration) and SV (co-rotation configuration and weaker SN feedback) and the gas-rich one (SIII-f50; see Table~\ref{table1}).
            The colours denote the different stage of interaction: before the first pericentre (blue), before the apocentre (green) and before the second pericentre (red). The vertical lines denote the times of the first pericentre (dotted lines), the apocentre ( dotted-dashed lines) and the second pericentre (dashed lines).
            For comparison, in the left panels we included the observational results from \citet[isolated galaxies: open squares, mergers: filled squares]{rupke2010}, \citet[black, open triangles]{queyrel2012}, \citet[black, open rombus]{jones2013}, \citet[black, inverted triangles]{jones2015}, \citet[black, crosses]{stott2014} and the linear regression reported by \citet[solid line]{stott2014}.}
            \label{slopessfrMM}
        \end{figure*}

        In Fig.~\ref{slopessfrMM} we show the S(O/H)-sSFR relation calculated for our four merger events: three of them correspond to the gas-poor initial condition run with different orientation parameters and SN feedback and the fourth one shows the relation for the gas-rich merger.
        The different stages of the merger evolution are depicted in different colours.
        As can be seen, all of them start away from the S(O/H)-sSFR relation.
        In the case of the gas-poor encounters, the S(O/H) does not change although the gas is being consumed into stars.
        Only when the galaxies are closer than $\sim 20$ kpc, there is an larger increase in the sSFR and a sudden flattening of the metallicity slopes.
        This occurs at about $0.1 - 0.15$ Gyr after the first passage.
        As a consequence, galaxies start defining a correlation that resembles closely the observed one reported by \citet{stott2014}.
        This behaviour is very similar to that of run SVI, where the interacting galaxies are counter-rotating. 

        The trend is not strongly affected by the strength of the SN feedback as shown by run SV which has five times lower SN energy per event.
        This is due to the combination of the deep potential well of the simulated systems and the quiescent star formation activity produced by the low gas fraction and the injection of SN energy in the ISM \citep{scan08}.
        However, a weaker SN feedback leads to a higher star formation activity since there is less gas heated up or ejected wind. 

        For the gas-rich encounter, the sSFR is larger during the first stages due to the larger gas fraction available in the system, even when the galaxies are far away (i.e. distance between the centre of masses larger than $\sim 100$ kpc).
        This star formation activity is fed by internal instabilities which formed clumps \citep{perez2011}.
        The triggering of the SFR is produced all over the disc where clumps are formed, leading to a more negative slope since the gas density is higher in the inner regions and star formation scheme is directly linked to the gas density.
        Only when the two galaxies are very close, the quick change of slope towards positive values set the galaxies well on the observed relation.

        In all runs, the on-set of the relation is clearly linked to the sudden change of metallicity slopes and sSFR associated to the close encounter at the same time.
        Previous works on galaxy-galaxy interactions \citep[e.g.][]{bh96,tissera2000,rupke2010} have shown the efficiency of these mechanisms to drive gas inflows, which transport low metallicity material to the centre, diluting the metallicities in the inner regions and triggering starbursts.
        This process has also taken place in our simulations as discussed in detailed in \citet{perez2011}. 
        Here, we go further in the analysis of the effects of gas inflows on both the metallicity gradients and the star formation activity.

    \subsection{Fly-by event and isolated galaxies}

\begin{figure*}
            \resizebox{17.7cm}{!}{\includegraphics{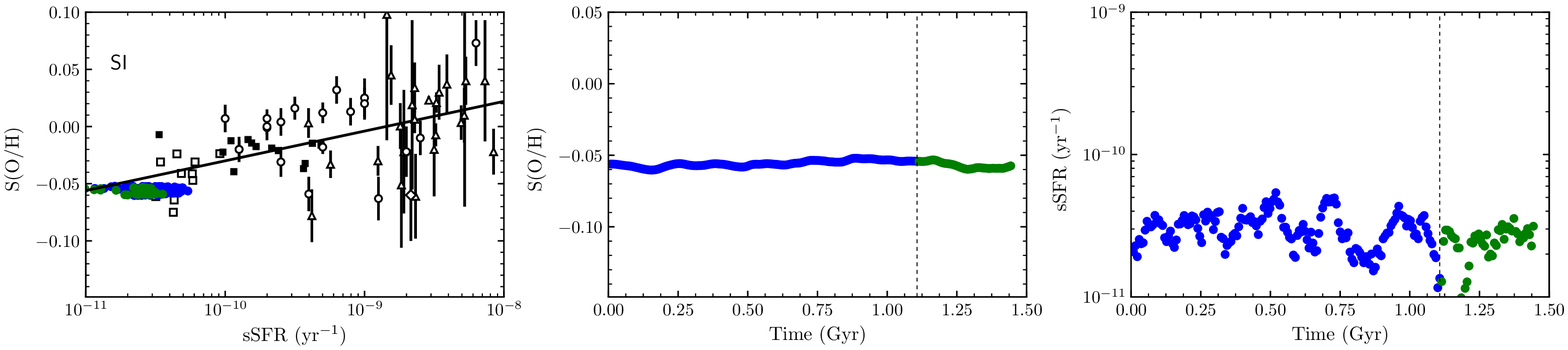}}
            \caption{Mean slope of the gas-phase oxygen abundance profiles as a function of the sSFR (left panels) and both quantities as a function of time (middle panels: gas phase oxygen slope; right panels: sSFR) for the simulated gas-phase for galaxies in a fly-by event (see Table~\ref{table1}). 
            For comparison, in the left panel we include the observational results as given in Fig.~\ref{slopessfrMM}.}
            \label{slopesfly}
        \end{figure*}

        Figure~\ref{slopesfly} shows the S(O/H)-sSFR relation for one of the galaxy pair in the major fly-by event (see Table~\ref{table1}). 
        In this galaxy, the metallicity gradient is not significantly modified along the interaction while the sSFR shows some variations.
        On average the sSFR varies mildly with time.
        This suggests that fly-by events are not able of developing strong inflows capable of increasing the star formation activity and modifying the metallicity slopes at the same time. 

\begin{figure*}
            \resizebox{17.7cm}{!}{\includegraphics{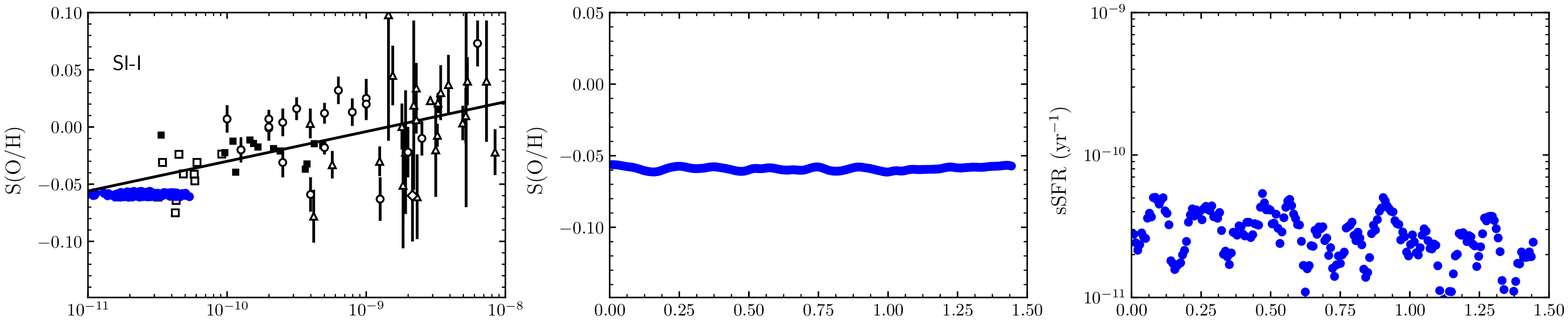}}
            \vspace*{-0.2cm}
            \resizebox{17.7cm}{!}{\includegraphics{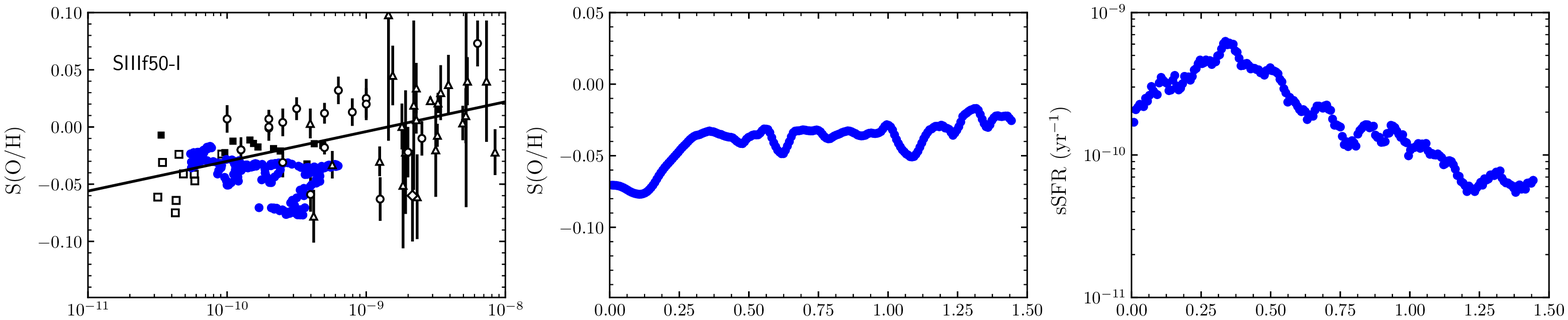}}
            \caption{Mean slope of the gas-phase oxygen abundance profiles as a function of the sSFR (left panels) and both quantities as a function of time (middle panels: gas phase oxygen slope; right panels: sSFR) for the simulated gas-phase for the gas poor (upper) and the gas-rich (lower) isolated galaxies (see \ref{table1}).
            For comparison, in the left panels we include the observational results as given in Fig.~\ref{slopessfrMM}.}
            \label{slopesiso}
        \end{figure*}

        Similarly, Fig.~\ref{slopesiso} shows the S(O/H)-sSFR relation for the isolated gas-poor (upper panels) and the isolated gas-rich (lower panels) systems.
        For gas-poor galaxy, the evolution of this relation is very similar to the fly-by event.

        However, the gas-rich galaxy is able to produce a significant change in both the metallicity slope and the sSFR.
        The regulation of the star formation activity in the gas-rich galaxy is produced by the formation of the clumps and the action of the SN feedback as it occurred in the first stages of the SIII-f50-I.
        The effects of both processes was discussed in details by \citet{perez2013}. 
        In this simulation, after the gas is consumed the star formation activity declines and there is no other mechanisms which can trigger gas inflows to feed a new starbursts as in the case of the galaxy interactions. 
        As the clumps migrate to the central region, the new stars form from a more chemically-enriched ISM.
        As a consequence, the metallicity gradient becomes slightly flatter and increases its zero point.
        Hence the S(O/H)-sSFR relation is only reached at the final stages.
        By comparing the evolution of the sSFR in SIII-f50 and SIII-f50-I, we can see that the interactions produce an important effect in the modulation of the star formation activity.
        We know that the interactions are able to redistribute the gas-phase component and to produce gas inflows which feed the central star formation activity.
        These runs suggest that the effects are more global in the sense that as the gas is shocked and tunnelled into to central regions, it is prevented to form stars, delaying the star formation activity until it is in condition to form stars again.
        Otherwise, the cold gas is transformed into stars very efficiently in clumps.
        Stronger SN feedback would help to modulate this process in a similar fashion \citep{perez2013}.
        The important aspect to consider is that both simulations (SIII-f50 and SIII-f50-I) have been run with the same initial conditions and the same SN feedback parameters.
        Hence the differences in their evolution relays on the presence of a companion.
        It is common to think that interactions are efficient to trigger star formation activity but they might also be important to modulate it during the long-range approach, by mixing and disturbing the gas-phase components, which is not longer in condition to form stars.

    \subsection{Gas inflows}

        The previous analysis suggests that if a correlation exists between S(O/H) and sSFR, it is more likely to be established during a period of strong star formation activity triggered by violent gas inflows.
        In order to quantify this scenario, we study the gas inflows, the metallicity of the infalling gas and the metallicity of the gas-phase component as a function of time for three examples: the SIII, SI and SI-I.
        Having in mind that these three simulations describe the evolution of the same galaxy in two different types of interaction (i.e. major merger and fly-by) and in isolation, these three runs highlight the effects of the presence of companion on the gas evolution.

        Gas inflows are estimated as the new gas mass identified within concentric radial shells at a given time.
        We define the ratio between the gas mass in the inflows and the gas mass in the radial shell before the inflow ($M_{\rm inflows}/M_{\rm bin}$).
        Note that as the galaxies approach each other, two processes can take place: contraction without shell-crossing or gas inflows producing shell-crossing.
        During the former, we expect an increase of the gas inflows in all the shells, in a similar fashion, while a genuine gas inflows will produce changes in selected shells, principally, the inner ones \citep{bh96}.

\begin{figure*}
            \resizebox{17.7cm}{!}{\includegraphics{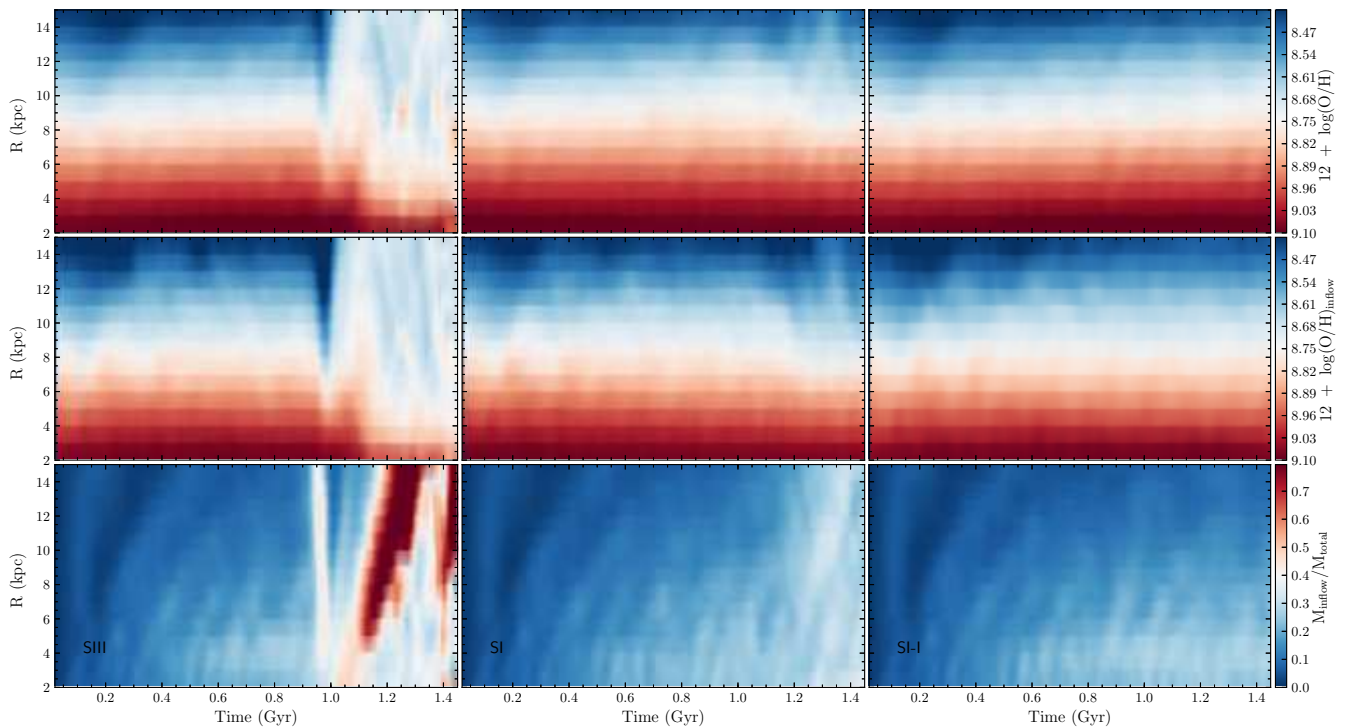}}\\
            \caption{Time evolution of median 12 + log(O/H) (upper panels ), the median metallicity of the infalling gas (middle panels) and the mass fraction of gas inflows (lower panels) estimated in concentric radial shells for runs SIII (left panels), SI (middle panels) and SI-I(right panels).}
            \label{inflows}
        \end{figure*}

        In Fig.~\ref{inflows} we show the metallicity of the gas in a given radial shell (upper panel), the metallicity of the infalling gas (middle panels) and the gas-mass fraction participating in the inflow (lower panels) for the SIII (first column), SI (second column) and SI-I (third column).

        In the first stages of evolution until the first pericentre, there are very weak changes in the metallicity gas distributions. 
        In the case of SIII, we can see the presence of gas inflows nearby the pericentre which transport lower metallicity gas, diluting the chemical abundances in the central regions as expected.
        The change in the slopes and sSFR is directly related to this low-metallicity gas inflows.

        On the other hand, in SI there is only a weak inflow after the close passage (the closer distance between the centre of masses is around 1.1 Gyr).
        In SI-I there are no important gas inflows, only a change in the global distribution of gas which gets slightly more concentrated without shell-crossing.
        From this figure it is clear that in order to change significantly the metallicity slope, a gas inflow representing more than $\sim~30 - 35$ per cent of the mass in the radial shell is needed.
        And that such inflow must be sudden and brief.
        Otherwise the gas feeds the star formation activity smoothly over time and the new enriched material will be smoothly incorporated to ISM, producing no correlated changes between the involved parameters.
        According to our simulations, violent inflows are mainly produced during mergers.
        In our simulations, fly-bys and isolated galaxies are not efficient mechanisms to do this. 
        Gas-rich disc galaxies might be able to produce the same effects via the formation of clumps as we show below. 
        In these simulations gas inflows fed by filaments are not considered. 
        However, they have been shown to be efficient at transporting low-metallicity gas into the galaxies \citep{ceverino2016} that might help the formation of clumps \citep[][and references therein]{tacchella2016}.

\begin{figure*}
            \resizebox{17.7cm}{!}{\includegraphics{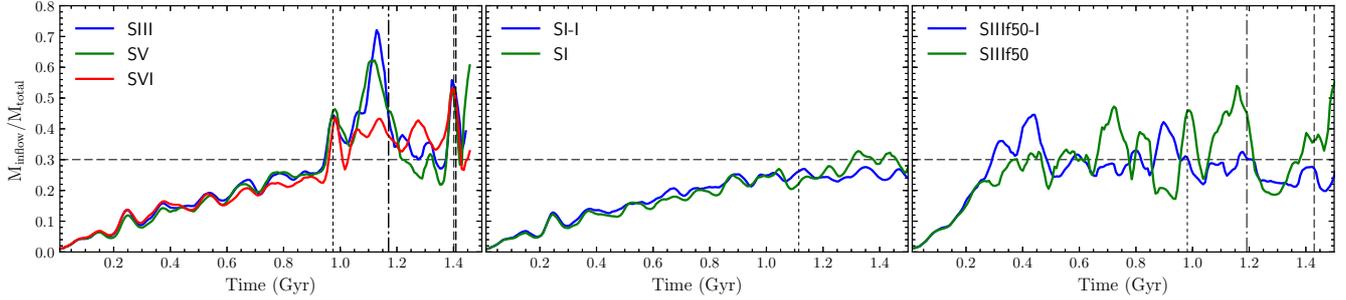}}
            \caption{Time evolution of median 12 + log(O/H) and the gas inflows onto the central region (within 5 kpc) for all simulations.
                The dotted lines represent the pericentre and the dotted-dashed lines the apocentre positions.}
            \label{inflows_5kp}
        \end{figure*}

        We estimate the fraction of gas mass associated to the inflows as a function of time for all our runs.
        We found that the systems which get to the S(O/H)-sSFR relation are those that experience gas inflows larger than $\sim 30-35$ per cent in short timescales, less than $\sim 0.2$ Gyr.
        Most of the strong inflows occurred within approximately $\sim 0.1$ Gyr.
        We define the inflow rate as the ration between the infalling gas within the central 5 kpc and the time interval over which it occurs, $(M_{in}/t_{in})$. 
        Then the inflow efficiency are estimated by dividing by the gas mass already present with the central region, $M_{tot}$.
        Our findings suggest that at least an inflow efficiency of $\sim 70$ Gyr$^{-1}$ is required to produce a simultaneous change in metallicity gradients and star formation rates. 
        This can be clearly seen from Fig.~\ref{inflows_5kp} where we show the gas inflows onto 5 kpc central region of all primary galaxies of our simulations.
        The gas inflows for the fly-by runs extend over larger periods, almost as continuous infall, producing no important change in the metallicity profile.
        As a consequence, the infalling gas has time to be transformed into stars and to smoothly enrich the remaining material.
        Only at $\sim 1.3$ Gyr there is a feeble peak in the inflows which last less than $\sim 0.1$ Gyr and is larger than 25 per cent which can be associated to a weak increase in the sSFR (see Fig.~\ref{slopesfly}).
        In the case of the isolated run of the same galaxy, there is a smooth increase over more than $\sim 1$ Gyr, producing an increase in the metallicity without modifying the metallicity gradient.

        These simulations illustrate clearly our findings: an important low-metallicity gas inflow occurring over a short time can set a galaxy on the S(O/H)-sSFR relation.
        If there is a larger gas fraction in the discs then internal instabilities might also set the galaxy on the relation but they have to be very efficient to produce the change in both the sSFR and the slope of the metallicity profiles at the same time.
        Extended mild SFR activity over time and over the whole discs is not efficient enough to produce a change in the metallicity slopes as indicated by our simulations.
        In this case the S(O/H) are preserved and the galaxies moved horizontally in the S(O/H)-sSFR plane.

        Finally we would like to note, that in the case of the galaxy-galaxy mergers, after the second pericentre, the outer regions of the discs might get contributions from the inner regions as the galaxy structure distorted (e.g while spiral arms are opened and deformed). 
        During these stages, there might also be mass exchange between the two galaxies that produces signals of gas inflows in outskirts of the galaxies.
        These inflows do not reach the central regions but they contribute with mildly enriched material in the external regions.
        These regions might also receive contributions from accreted gas via galactic fountain as detected by \citet{perez2011}.
        These effects can be separated from those related to inner gas inflows, if mass intervals are used instead of radial shells.
        By doing so, variations due to coherent re-arrangement of material (compression or expansion) are not considered and only those which produce shell mixing are unveiled.
        We have done these estimations finding trends which are in agreement with the results presented here.
        We choose to show radial shells in order to compare with the results from previous works \citep{perez2011, perez2013}.
        In the following section, we will analyse the effects on incorporating the outskirts of interacting discs in the determination of the metallicity profiles.

    \subsection{The metallicity gradients in the outer parts}

\begin{figure*}
        	\subfloat{\includegraphics[width=8.5cm]{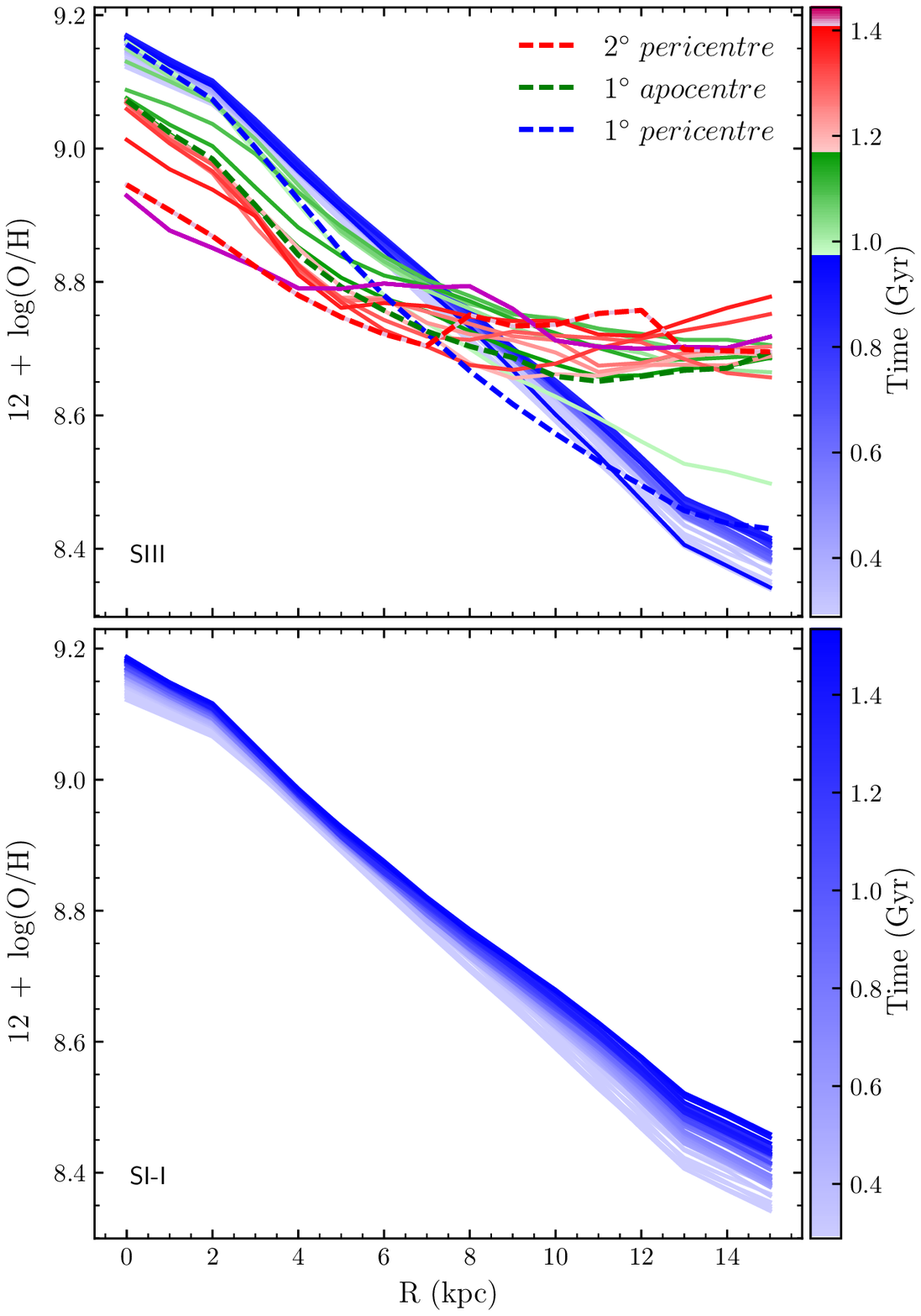}}
        	\subfloat{\includegraphics[width=8.5cm]{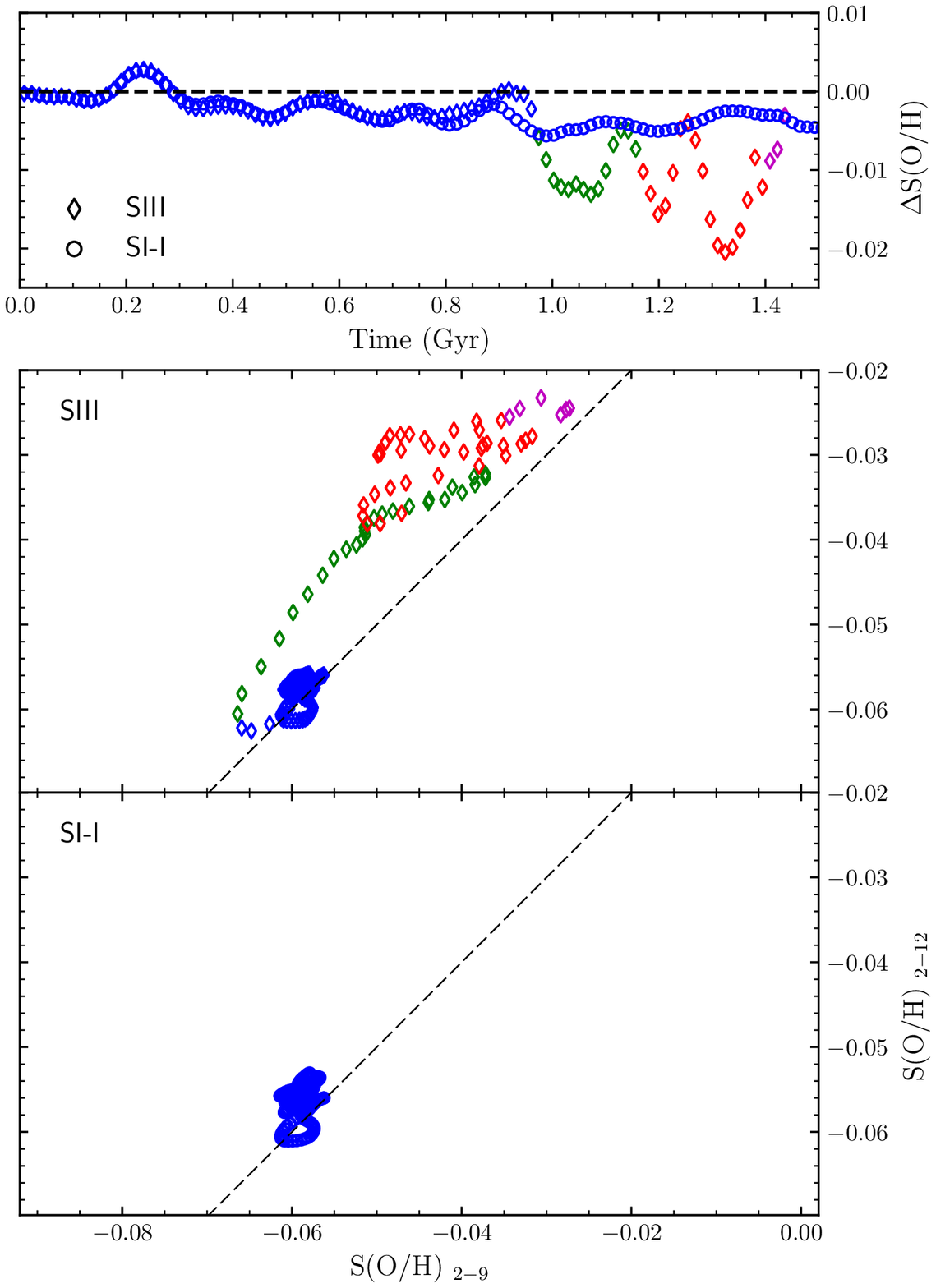}}
            \caption{Left panel: Time evolution of the metallicity gradients for the SIII event (upper panel) and the isolated counterpart (SI-I, lower panel). 
            Right panel: Slope of the metallicity gradients estimated in the range $[2 - 9]$ kpc versus those measured in the range $[2 - 12]$ kpc for the SIII simulations and and SI-I (middle and right lower panels) 
            The upper left panel shows the relative differences between them.
            The different colours denote different stages of evolution during the interaction (see Fig.1).}
            \label{slopediff}
        \end{figure*}

        It is well known that, at $z \sim 0$, the outskirts of the disc galaxies have the metallicity gradients which do not follow the extrapolated linear relation of the inner parts \citep[e.g.][]{sanchez-menguiano2016}.
        This departure from a linear profile in the sense that outer parts are more gas-enriched than expected, is also reproduced in the cosmological simulations of T2016 and in the pre-prepared merger events \citet{perez2011}. 
        The latter work also showed how the outer parts of galaxies are more easily disturbed during the interaction while fed by recycled material.
        In Fig.~\ref{slopediff} (left panel) we show the metallicity gradients as a function of time for runs SIII and SI-I, for illustration purposes.
        This figure shows clearly the difference between the evolution of the gas-phase metallicity gradients in both situations.
        We recall that the initial conditions of each galaxy of these simulations are the same, differing only in the fact that SIII system describes a merger event of two equal-mass galaxies and SI-I follows the evolution of one of these galaxies in isolation. 
        A break in the metallicity profiles originates between the first and the second pericentre and is produced by both the spiral arms which get distorted and the accreted material that is ejected after the starbursts triggered during the first approach.
        In the case of the isolated galaxy, there is a monotonically increase of the metallicity at all radii as new stars formed without modifying the global profile.
        The flattening of the metallicity profiles in the outskirts of galaxies have been reported by observations of spiral galaxies \citep[e.g.][and references therein]{sanchez-menguiano2016}. 
        Our simulations show how the exchange of material by the two mentioned physical mechanisms can explain the break of the metallicity profiles in the outskirts of disc galaxies

        To test the impact of including the outer parts in the estimation of the metallicity gradients, in Fig.~\ref{slopediff} (right panel) we compare the S(O/H) obtained from the linear regression performed within two different radial ranges: $[2 - 9]$ kpc and $[2 - 12]$ kpc.
        In this figure (right bottom panel), we plot the $[2 - 9]$ kpc slope against the $[2 - 12]$ kpc one.
        As can be appreciated, the gradients are quite similar for SIII and SI-I before the first close passage.
        After that the SIII shows steeper metallicity gradients in the inner regions ($[2 - 9]$ kpc) along the whole interaction until the merger.
        When the galaxies are very close (after the apocentre) the flattening of the metallicity gradient estimated in the range $[2 - 12]$ kpc is clear.
        In Fig.~\ref{slopediff} (left upper panel) we show the relative change between the slopes. 
        The differences of the slopes in SIII are produced by the joint effects of the opening of spiral arms and the recycled material via galactic fountain.
        Hence, it is very important to be aware of the impact of these effects if these regions are included in the estimations of the metallicity profiles.
        This might be an important issue to consider for the determination of metallicity gradients, particularly at high redshift where galaxies might be even more affected by gas inflows and mergers.
        It is also important to note that these pre-prepared merger events are not enough to understand the complexity of the origin and evolution of the metallicity gradients in a cosmological context but they give a clearer description of the effects of individual processes.
        
\section{Discussion: Comparison with cosmological simulation}

    The analysis presented above involves a set of hydrodynamical simulations of isolated and interacting pre-prepared disc galaxies of similar masses. 
    Hence it is interesting to compare them with those obtained from galaxies that formed in a full cosmological framework.
    For this purpose, we resort to the numerical results reported by T2016 who studied the chemical abundance profiles of the gas-phase disc components in relation to the star formation activity of the galaxies simulated in a hierarchical universe consistent with a $\Lambda$CDM scenario.
    The cosmological simulation takes into account other environmental effects such as gas infall along filaments or minor accretions which are not considered by the pre-prepared initial conditions analysed in the previous sections. 
    These processes can affect the metallicity gradients by redistributing chemical elements in the gaseous phase \citep[e.g.][]{tissera2016a,ma2017}.
    The comparison with a cosmological simulation is done to assess at what extend the effects of the galaxy-galaxy interactions can still be present in a more complex, scenario for galaxy formation. 
    This cosmological simulation has been reported to produce galaxies with sizes and angular momentum content comparable to observations as shown in \citet{pedrosa2015}. 
    The chemical abundances and star formation activity of the disc components are analysed in detail by T2016.
    These authors report that the simulated metallicity gradients are in agreement with the observational trends between the slopes of the metallicity profiles and the galaxy stellar masses \citep{ho2015} and between the metallicity slopes and the sSFRs \citep{stott2014}.
    The cosmological simulation of T2016 and our pre-prepared simulations were run with the same version of P-Gadget-3 which used the same SN feedback model (i.e. energy and chemical feedback).
    The numerical resolution of the simulations are similar (slightly better in the pre-prepared merger).
    Hence, the comparison between the metallicity gradients can be done on equal bases. 
    The pre-prepared mergers allow us to better separate the effects of interactions and the presence of companions. Although cosmological simulations provide more realistic assembly histories, the non-linear evolution of the structure makes it more difficult to disentangle the action of different physical mechanisms.
    Hence the confrontation of results from both types of simulations opens the possibility to get a further insight into this relation.

\begin{figure}
        \resizebox{8.5cm}{!}{\includegraphics{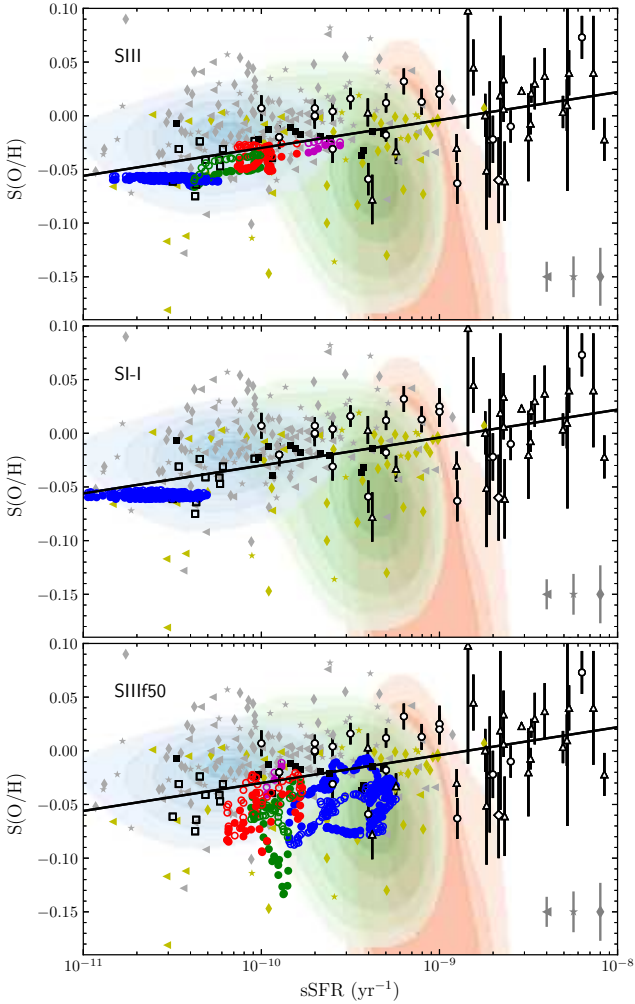}}\\
        \caption{Comparison of the metallicity slopes and sSFR obtained from a $\Lambda$-CDM hydrodynamical simulation \citep{tissera2016a} and those measured in the pre-prepared simulations: SIII (upper panel), SI-I (middle panel) and SIII-f50 (lower panel). 
        The light blue (green, red) contours enclose 90, 70, 50, 30 and 20 per cent of the simulated disc galaxies at $z \sim 0 \ (1,2)$ in the cosmological simulation.
        The black symbols represent observational data (see Fig.1 for details) and grey and yellow symbols are the observational data obtained by \citet{wuyts2016} and the simulated sample performed by \citet{ma2017} respectively ($z\sim 0$: triangle; $z\sim 1$: stars; $z\sim 2$: diamond).
	    The error bars in the right bottom corner of each panel represent the median of the errors by redshift interval.}
        \label{slopecosmo}
    \end{figure}

    In Fig~\ref{slopecosmo}, the results from T2016 are shown as the light blue (green, red) contours at $z \sim 0\ (1, 2)$ which enclose 90, 70, 50, 30 and 20 per cent of the galaxies. 
    The results for SIII, SI, and SIII-f50 are also displayed in this figure (upper, middle and lower panels).
    The gas-rich simulation has the larger dispersion as shown in Fig.~\ref{slopesiso} although the values are within the range predicted by the cosmological simulation and obtained from the available observations. 
    In the case of galaxies simulated in a cosmological framework, they all have experienced mergers and interactions.
    The result of this evolutionary scenario is the formation of disc galaxies whose gas-phase components show metallicity gradients in global agreement with observations.

    As reported in the previous sections, it is also clear that negative metallicity gradients can be generated during gas-rich mergers that are likely responsible of the negative gradients detected in the cosmological simulations at higher redshift (T2016). 
    From an observational point of view, metallicity gradients are measured as a function of redshift and showing a large dispersion with no clear trend \citep[e.g.][]{wuyts2016}.
    We also know few cases of negative metallicity gradients at $z > 1$ obtained through gravitational lensing \citep{yuan2011,jones2013,jones2015,leethochawalit2016}.
    Although the determination of global metallicities and metallicity gradients are affected by uncertainties which make it difficult to draw a robust conclusions, there is a increasing database of metallicity gradients available for comparison with models.
    In Fig~\ref{slopecosmo} we show the observational data gathered up to $z \sim 2.6$, which yields a Pearson correlation factor of $r = 0.39$ ($p = 5.4 \times 10^{-8}$). 
    For the observational data of \citet{wuyts2016} we obtain $r = 0.33$ ($p = 5.4 \times 10^{-4}$) and for the rest of the data $r = 0.49$ ($p = 1.7 \times 10^{-5}$).
    Although metallicity gradients at high redshift have larger error bars, the statistical signal is significant.

    Our results predict a correlation between the metallicity gradients and the sSFR produced by the effects of violent inflows which link together their evolution. 
    The large dispersion found in this correlation reflects the variation in galaxy properties and their environment as a function of redshift as can be seen from the trends reported by T2016 in a cosmological context.
    Moreover, similar trends have been also found by \citet{ma2017} using a set of simulated zoom-in galaxies run with a different code and SN feedback model (Fig~\ref{slopecosmo})

\section{Conclusions}

    Gas inflows produced during galaxy-galaxy interactions or as a result of strong internal instabilities can produce both the rapid dilution of the central metallicity of the gas-phase and the quick increase of the sSFR, setting a correlation between metallicity gradients and sSFR \citet{stott2014}.
    Our findings suggest that the correlation is established when galaxy-galaxy interactions are efficient mechanism to trigger gas inflows involving more than 35 per cent of the existing gas mass in the central regions during a time interval less than $\sim 5 \times 10^8$ yr.

    We also found that isolated gas-rich galaxies can set a similar correlation between the metallicity gradients and the sSFR as a consequence of the formation of clumps where the star formation activity is efficiently triggered.
    For our gas-rich simulation, \citet{perez2013} determined that one per cent of the clump mass per free-fall time was converted into stars, producing bound systems with high probability of survival \citet{krumholz2010}.
    The formation of clumps is proportional to the gas density and they migrate from the outer into to the inner parts carrying lower metallicity material).
    This process could be also contribute to set a correlation between the metallicity gradients and sSFR, principally at high redshift, where gas-rich discs are more frequent \citep{ceverino2016, molina2016}.

    A galaxy with approximately uniform star formation activity will consume its gas reservoir in a more regular fashion, producing an extended star formation activity and a slow enrichment of the gas-phase which does not change the metallicity gradient significantly.
    Such system would not evolve according to this relation but move horizontally as the sSFR decreases with a smaller dispersion than in the cases of the gas-rich simulations.
    Nevertheless, the simulated relations are within observed range of parameters.

    The comparison with the cosmological results from T2016 are in good agreement showing that a hierarchical clustering scenario provide favourable conditions for the existence of the S(O/H)-sSFR correlation.
    With the increase of database of galaxies with well-determined chemical abundances in the local and high-z \citep{wuyts2016}, such a correlation might indeed be present, albeit with a large dispersion.
    The origin of the dispersion may reside in the combination of physical mechanisms triggering star formation in either violent or smooth fashions.

\section*{Acknowledgements}

    We acknowledged the use of Fenix Cluster of Institute for Astronomy and Space Physics (Argentina).
    This work has been partially supported by PICT Raices 2011/959 of Ministry of Science (Argentina), Proyecto Interno 677-15/N - UNAB (Chile), Project Fondecyt 1150334 (Chile) and the Network SAN supported by Redes 150078 Conicyt (Chile).

\bibliography{PairsMet}{}

\begin{thebibliography}{}
\makeatletter
\relax
\def\mn@urlcharsother{\let\do\@makeother \do\$\do\&\do\#\do\^\do\_\do\%\do\~}
\def\mn@doi{\begingroup\mn@urlcharsother \@ifnextchar [ {\mn@doi@}
  {\mn@doi@[]}}
\def\mn@doi@[#1]#2{\def\@tempa{#1}\ifx\@tempa\@empty \href
  {http://dx.doi.org/#2} {doi:#2}\else \href {http://dx.doi.org/#2} {#1}\fi
  \endgroup}
\def\mn@eprint#1#2{\mn@eprint@#1:#2::\@nil}
\def\mn@eprint@arXiv#1{\href {http://arxiv.org/abs/#1} {{\tt arXiv:#1}}}
\def\mn@eprint@dblp#1{\href {http://dblp.uni-trier.de/rec/bibtex/#1.xml}
  {dblp:#1}}
\def\mn@eprint@#1:#2:#3:#4\@nil{\def\@tempa {#1}\def\@tempb {#2}\def\@tempc
  {#3}\ifx \@tempc \@empty \let \@tempc \@tempb \let \@tempb \@tempa \fi \ifx
  \@tempb \@empty \def\@tempb {arXiv}\fi \@ifundefined
  {mn@eprint@\@tempb}{\@tempb:\@tempc}{\expandafter \expandafter \csname
  mn@eprint@\@tempb\endcsname \expandafter{\@tempc}}}

\bibitem[\protect\citeauthoryear{{Angl{\'e}s-Alc{\'a}zar}, {Dav{\'e}},
  {{\"O}zel}  \& {Oppenheimer}}{{Angl{\'e}s-Alc{\'a}zar}
  et~al.}{2014}]{angles-alcazar2014}
{Angl{\'e}s-Alc{\'a}zar} D.,  {Dav{\'e}} R.,  {{\"O}zel} F.,   {Oppenheimer}
  B.~D.,  2014, \mn@doi [\apj] {10.1088/0004-637X/782/2/84}, \href
  {http://adsabs.harvard.edu/abs/2014ApJ...782...84A} {782, 84}

\bibitem[\protect\citeauthoryear{{Barnes} \& {Hernquist}}{{Barnes} \&
  {Hernquist}}{1996}]{bh96}
{Barnes} J.~E.,  {Hernquist} L.,  1996, \mn@doi [ApJ] {10.1086/177957}, \href
  {http://adsabs.harvard.edu/abs/1996ApJ...471..115B} {471, 115}

\bibitem[\protect\citeauthoryear{{Barton}, {Geller}  \& {Kenyon}}{{Barton}
  et~al.}{2000}]{barton2000}
{Barton} E.~J.,  {Geller} M.~J.,   {Kenyon} S.~J.,  2000, \mn@doi [\apj]
  {10.1086/308392}, \href {http://adsabs.harvard.edu/abs/2000ApJ...530..660B}
  {530, 660}

\bibitem[\protect\citeauthoryear{{Belfiore} et~al.,}{{Belfiore}
  et~al.}{2017}]{belfiore2017}
{Belfiore} F.,  et~al., 2017, preprint, \href
  {http://adsabs.harvard.edu/abs/2017arXiv170303813B} {} (\mn@eprint {arXiv}
  {1703.03813})

\bibitem[\protect\citeauthoryear{{Bournaud} et~al.,}{{Bournaud}
  et~al.}{2011}]{bournaud2011}
{Bournaud} F.,  et~al., 2011, \mn@doi [\apj] {10.1088/0004-637X/730/1/4}, \href
  {http://adsabs.harvard.edu/abs/2011ApJ...730....4B} {730, 4}

\bibitem[\protect\citeauthoryear{{Calura} et~al.,}{{Calura}
  et~al.}{2012}]{calura2012}
{Calura} F.,  et~al., 2012, \mn@doi [MNRAS] {10.1111/j.1365-2966.2012.22052.x},
  \href {http://adsabs.harvard.edu/abs/2012MNRAS.427.1401C} {427, 1401}

\bibitem[\protect\citeauthoryear{{Ceverino}, {S{\'a}nchez Almeida}, {Mu{\~n}oz
  Tu{\~n}{\'o}n}, {Dekel}, {Elmegreen}, {Elmegreen}  \& {Primack}}{{Ceverino}
  et~al.}{2016}]{ceverino2016}
{Ceverino} D.,  {S{\'a}nchez Almeida} J.,  {Mu{\~n}oz Tu{\~n}{\'o}n} C.,
  {Dekel} A.,  {Elmegreen} B.~G.,  {Elmegreen} D.~M.,   {Primack} J.,  2016,
  \mn@doi [\mnras] {10.1093/mnras/stw064}, \href
  {http://adsabs.harvard.edu/abs/2016MNRAS.457.2605C} {457, 2605}

\bibitem[\protect\citeauthoryear{{Chiappini}, {Matteucci}  \&
  {Gratton}}{{Chiappini} et~al.}{1997}]{chiap1997}
{Chiappini} C.,  {Matteucci} F.,   {Gratton} R.,  1997, \mn@doi [ApJ]
  {10.1086/303726}, \href {http://adsabs.harvard.edu/abs/1997ApJ...477..765C}
  {477, 765}

\bibitem[\protect\citeauthoryear{{Crain} et~al.,}{{Crain}
  et~al.}{2015}]{crain2015}
{Crain} R.~A.,  et~al., 2015, \mn@doi [\mnras] {10.1093/mnras/stv725}, \href
  {http://adsabs.harvard.edu/abs/2015MNRAS.450.1937C} {450, 1937}

\bibitem[\protect\citeauthoryear{{Dutil} \& {Roy}}{{Dutil} \&
  {Roy}}{1999}]{dutil1999}
{Dutil} Y.,  {Roy} J.,  1999, \mn@doi [ApJ] {10.1086/307100}, \href
  {http://adsabs.harvard.edu/abs/1999ApJ...516...62D} {516, 62}

\bibitem[\protect\citeauthoryear{{Ellison}, {Patton}, {Simard}, {McConnachie},
  {Baldry}  \& {Mendel}}{{Ellison} et~al.}{2010}]{ellison2010}
{Ellison} S.~L.,  {Patton} D.~R.,  {Simard} L.,  {McConnachie} A.~W.,  {Baldry}
  I.~K.,   {Mendel} J.~T.,  2010, \mn@doi [\mnras]
  {10.1111/j.1365-2966.2010.17076.x}, \href
  {http://adsabs.harvard.edu/abs/2010MNRAS.407.1514E} {407, 1514}

\bibitem[\protect\citeauthoryear{{Gibson}, {Pilkington}, {Brook}, {Stinson}  \&
  {Bailin}}{{Gibson} et~al.}{2013}]{gibson2013}
{Gibson} B.~K.,  {Pilkington} K.,  {Brook} C.~B.,  {Stinson} G.~S.,   {Bailin}
  J.,  2013, \mn@doi [A\&A] {10.1051/0004-6361/201321239}, \href
  {http://adsabs.harvard.edu/abs/2013A%26A...554A..47G} {554, A47}

\bibitem[\protect\citeauthoryear{{Ho}, {Kudritzki}, {Kewley}, {Zahid},
  {Dopita}, {Bresolin}  \& {Rupke}}{{Ho} et~al.}{2015}]{ho2015}
{Ho} I.-T.,  {Kudritzki} R.-P.,  {Kewley} L.~J.,  {Zahid} H.~J.,  {Dopita}
  M.~A.,  {Bresolin} F.,   {Rupke} D.~S.~N.,  2015, \mn@doi [MNRAS]
  {10.1093/mnras/stv067}, \href
  {http://adsabs.harvard.edu/abs/2015MNRAS.448.2030H} {448, 2030}

\bibitem[\protect\citeauthoryear{{Iwamoto}, {Brachwitz}, {Nomoto}, {Kishimoto},
  {Umeda}, {Hix}  \& {Thielemann}}{{Iwamoto} et~al.}{1999}]{iwamoto1999}
{Iwamoto} K.,  {Brachwitz} F.,  {Nomoto} K.,  {Kishimoto} N.,  {Umeda} H.,
  {Hix} W.~R.,   {Thielemann} F.-K.,  1999, \mn@doi [ApJS] {10.1086/313278},
  \href {http://adsabs.harvard.edu/abs/1999ApJS..125..439I} {125, 439}

\bibitem[\protect\citeauthoryear{{Jones}, {Ellis}, {Richard}  \&
  {Jullo}}{{Jones} et~al.}{2013}]{jones2013}
{Jones} T.,  {Ellis} R.~S.,  {Richard} J.,   {Jullo} E.,  2013, \mn@doi [ApJ]
  {10.1088/0004-637X/765/1/48}, \href
  {http://adsabs.harvard.edu/abs/2013ApJ...765...48J} {765, 48}

\bibitem[\protect\citeauthoryear{{Jones} et~al.,}{{Jones}
  et~al.}{2015}]{jones2015}
{Jones} T.,  et~al., 2015, \mn@doi [AJ] {10.1088/0004-6256/149/3/107}, \href
  {http://adsabs.harvard.edu/abs/2015AJ....149..107J} {149, 107}

\bibitem[\protect\citeauthoryear{{Kewley}, {Geller}  \& {Barton}}{{Kewley}
  et~al.}{2006}]{kewley2006}
{Kewley} L.~J.,  {Geller} M.~J.,   {Barton} E.~J.,  2006, \mn@doi [AJ]
  {10.1086/500295}, \href {http://adsabs.harvard.edu/abs/2006AJ....131.2004K}
  {131, 2004}

\bibitem[\protect\citeauthoryear{{Kewley}, {Rupke}, {Zahid}, {Geller}  \&
  {Barton}}{{Kewley} et~al.}{2010}]{kewley2010}
{Kewley} L.~J.,  {Rupke} D.,  {Zahid} H.~J.,  {Geller} M.~J.,   {Barton} E.~J.,
   2010, \mn@doi [ApJL] {10.1088/2041-8205/721/1/L48}, \href
  {http://adsabs.harvard.edu/abs/2010ApJ...721L..48K} {721, L48}

\bibitem[\protect\citeauthoryear{{Khochfar} \& {Burkert}}{{Khochfar} \&
  {Burkert}}{2006}]{khochfar2006}
{Khochfar} S.,  {Burkert} A.,  2006, \mn@doi [\aap]
  {10.1051/0004-6361:20053241}, \href
  {http://adsabs.harvard.edu/abs/2006A%26A...445..403K} {445, 403}

\bibitem[\protect\citeauthoryear{{Krumholz} \& {Dekel}}{{Krumholz} \&
  {Dekel}}{2010}]{krumholz2010}
{Krumholz} M.~R.,  {Dekel} A.,  2010, \mn@doi [\mnras]
  {10.1111/j.1365-2966.2010.16675.x}, \href
  {http://adsabs.harvard.edu/abs/2010MNRAS.406..112K} {406, 112}

\bibitem[\protect\citeauthoryear{{Lambas}, {Tissera}, {Alonso}  \&
  {Coldwell}}{{Lambas} et~al.}{2003}]{lambas2003}
{Lambas} D.~G.,  {Tissera} P.~B.,  {Alonso} M.~S.,   {Coldwell} G.,  2003,
  \mn@doi [\mnras] {10.1111/j.1365-2966.2003.07179.x}, \href
  {http://adsabs.harvard.edu/abs/2003MNRAS.346.1189L} {346, 1189}

\bibitem[\protect\citeauthoryear{{Leethochawalit}, {Jones}, {Ellis}, {Stark},
  {Richard}, {Zitrin}  \& {Auger}}{{Leethochawalit}
  et~al.}{2016}]{leethochawalit2016}
{Leethochawalit} N.,  {Jones} T.~A.,  {Ellis} R.~S.,  {Stark} D.~P.,  {Richard}
  J.,  {Zitrin} A.,   {Auger} M.,  2016, \mn@doi [\apj]
  {10.3847/0004-637X/820/2/84}, \href
  {http://adsabs.harvard.edu/abs/2016ApJ...820...84L} {820, 84}

\bibitem[\protect\citeauthoryear{{Lequeux}, {Peimbert}, {Rayo}, {Serrano}  \&
  {Torres-Peimbert}}{{Lequeux} et~al.}{1979}]{lequeux1979}
{Lequeux} J.,  {Peimbert} M.,  {Rayo} J.~F.,  {Serrano} A.,   {Torres-Peimbert}
  S.,  1979, A\&A, \href {http://adsabs.harvard.edu/abs/1979A%26A....80..155L}
  {80, 155}

\bibitem[\protect\citeauthoryear{{Ma}, {Hopkins}, {Wetzel}, {Kirby},
  {Angl{\'e}s-Alc{\'a}zar}, {Faucher-Gigu{\`e}re}, {Kere{\v s}}  \&
  {Quataert}}{{Ma} et~al.}{2017}]{ma2017}
{Ma} X.,  {Hopkins} P.~F.,  {Wetzel} A.~R.,  {Kirby} E.~N.,
  {Angl{\'e}s-Alc{\'a}zar} D.,  {Faucher-Gigu{\`e}re} C.-A.,  {Kere{\v s}} D.,
   {Quataert} E.,  2017, \mn@doi [\mnras] {10.1093/mnras/stx273}, \href
  {http://adsabs.harvard.edu/abs/2017MNRAS.467.2430M} {467, 2430}

\bibitem[\protect\citeauthoryear{{Madau} \& {Dickinson}}{{Madau} \&
  {Dickinson}}{2014}]{madau2014}
{Madau} P.,  {Dickinson} M.,  2014, \mn@doi [\araa]
  {10.1146/annurev-astro-081811-125615}, \href
  {http://adsabs.harvard.edu/abs/2014ARA%26A..52..415M} {52, 415}

\bibitem[\protect\citeauthoryear{{Maiolino}, {Nagao}, {Grazian}, {Cocchia},
  {Marconi}, {Mannucci}, {Cimatti}  \& {Pipino}}{{Maiolino}
  et~al.}{2008}]{maiolino2008}
{Maiolino} R.,  {Nagao} T.,  {Grazian} A.,  {Cocchia} F.,  {Marconi} A.,
  {Mannucci} F.,  {Cimatti} A.,   {Pipino} A. e.~a.,  2008, \mn@doi [A\&A]
  {10.1051/0004-6361:200809678}, \href
  {http://adsabs.harvard.edu/abs/2008A%26A...488..463M} {488, 463}

\bibitem[\protect\citeauthoryear{{Michel-Dansac}, {Lambas}, {Alonso}  \&
  {Tissera}}{{Michel-Dansac} et~al.}{2008}]{dansac2008}
{Michel-Dansac} L.,  {Lambas} D.~G.,  {Alonso} M.~S.,   {Tissera} P.,  2008,
  \mn@doi [MNRAS] {10.1111/j.1745-3933.2008.00466.x}, \href
  {http://adsabs.harvard.edu/abs/2008MNRAS.386L..82M} {386, L82}

\bibitem[\protect\citeauthoryear{{Mihos} \& {Hernquist}}{{Mihos} \&
  {Hernquist}}{1996}]{mh96}
{Mihos} J.~C.,  {Hernquist} L.,  1996, \mn@doi [ApJ] {10.1086/177353}, \href
  {http://adsabs.harvard.edu/abs/1996ApJ...464..641M} {464, 641}

\bibitem[\protect\citeauthoryear{{Molina}, {Ibar}, {Swinbank}, {Sobral},
  {Best}, {Smail}, {Escala}  \& {Cirasuolo}}{{Molina}
  et~al.}{2016}]{molina2016}
{Molina} J.,  {Ibar} E.,  {Swinbank} A.~M.,  {Sobral} D.,  {Best} P.~N.,
  {Smail} I.,  {Escala} A.,   {Cirasuolo} M.,  2016, preprint, \href
  {http://adsabs.harvard.edu/abs/2016arXiv161200447M} {} (\mn@eprint {arXiv}
  {1612.00447})

\bibitem[\protect\citeauthoryear{{Mosconi}, {Tissera}, {Lambas}  \&
  {Cora}}{{Mosconi} et~al.}{2001}]{mosconi2001}
{Mosconi} M.~B.,  {Tissera} P.~B.,  {Lambas} D.~G.,   {Cora} S.~A.,  2001,
  \mn@doi [MNRAS] {10.1046/j.1365-8711.2001.04198.x}, \href
  {http://adsabs.harvard.edu/abs/2001MNRAS.325...34M} {325, 34}

\bibitem[\protect\citeauthoryear{{Muratov} et~al.,}{{Muratov}
  et~al.}{2017}]{muratov2017}
{Muratov} A.~L.,  et~al., 2017, \mn@doi [\mnras] {10.1093/mnras/stx667}, \href
  {http://adsabs.harvard.edu/abs/2017MNRAS.468.4170M} {468, 4170}

\bibitem[\protect\citeauthoryear{{Navarro}, {Frenk}  \& {White}}{{Navarro}
  et~al.}{1996}]{nfw96}
{Navarro} J.~F.,  {Frenk} C.~S.,   {White} S.~D.~M.,  1996, \mn@doi [\apj]
  {10.1086/177173}, \href {http://adsabs.harvard.edu/abs/1996ApJ...462..563N}
  {462, 563}

\bibitem[\protect\citeauthoryear{{Patton}, {Ellison}, {Simard}, {McConnachie}
  \& {Mendel}}{{Patton} et~al.}{2011}]{patton2011}
{Patton} D.~R.,  {Ellison} S.~L.,  {Simard} L.,  {McConnachie} A.~W.,
  {Mendel} J.~T.,  2011, \mn@doi [\mnras] {10.1111/j.1365-2966.2010.17932.x},
  \href {http://adsabs.harvard.edu/abs/2011MNRAS.412..591P} {412, 591}

\bibitem[\protect\citeauthoryear{{Pedrosa} \& {Tissera}}{{Pedrosa} \&
  {Tissera}}{2015}]{pedrosa2015}
{Pedrosa} S.,  {Tissera} P.,  2015, preprint, \href
  {http://adsabs.harvard.edu/abs/2015arXiv150807220P} {} (\mn@eprint {arXiv}
  {1508.07220})

\bibitem[\protect\citeauthoryear{{P{\'e}rez-Montero}, {Garc{\'{\i}}a-Benito},
  {V{\'{\i}}lchez}, {S{\'a}nchez}, {Kehrig}, {Husemann}, ...  \& {Califa
  Collaboration}}{{P{\'e}rez-Montero} et~al.}{2016}]{perez-montero2016}
{P{\'e}rez-Montero} E.,  {Garc{\'{\i}}a-Benito} R.,  {V{\'{\i}}lchez} J.~M.,
  {S{\'a}nchez} S.~F.,  {Kehrig} C.,  {Husemann} B.,  ...  {Califa
  Collaboration} 2016, \mn@doi [\aap] {10.1051/0004-6361/201628601}, \href
  {http://adsabs.harvard.edu/abs/2016A%26A...595A..62P} {595, A62}

\bibitem[\protect\citeauthoryear{{Perez}, {Michel-Dansac}  \&
  {Tissera}}{{Perez} et~al.}{2011}]{perez2011}
{Perez} J.,  {Michel-Dansac} L.,   {Tissera} P.~B.,  2011, \mn@doi [MNRAS]
  {10.1111/j.1365-2966.2011.19300.x}, \href
  {http://adsabs.harvard.edu/abs/2011MNRAS.417..580P} {417, 580}

\bibitem[\protect\citeauthoryear{{Perez}, {Valenzuela}, {Tissera}  \&
  {Michel-Dansac}}{{Perez} et~al.}{2013}]{perez2013}
{Perez} J.,  {Valenzuela} O.,  {Tissera} P.~B.,   {Michel-Dansac} L.,  2013,
  \mn@doi [MNRAS] {10.1093/mnras/stt1563}, \href
  {http://adsabs.harvard.edu/abs/2013MNRAS.436..259P} {436, 259}

\bibitem[\protect\citeauthoryear{{Pilkington} et~al.,}{{Pilkington}
  et~al.}{2012}]{pilkington2012}
{Pilkington} K.,  et~al., 2012, \mn@doi [MNRAS]
  {10.1111/j.1365-2966.2012.21353.x}, \href
  {http://adsabs.harvard.edu/abs/2012MNRAS.425..969P} {425, 969}

\bibitem[\protect\citeauthoryear{{Queyrel} et~al.,}{{Queyrel}
  et~al.}{2012}]{queyrel2012}
{Queyrel} J.,  et~al., 2012, \mn@doi [A\&A] {10.1051/0004-6361/201117718},
  \href {http://adsabs.harvard.edu/abs/2012A%26A...539A..93Q} {539, A93}

\bibitem[\protect\citeauthoryear{{Rosas-Guevara} et~al.,}{{Rosas-Guevara}
  et~al.}{2015}]{rosas-guevara2015}
{Rosas-Guevara} Y.~M.,  et~al., 2015, \mn@doi [\mnras] {10.1093/mnras/stv2056},
  \href {http://adsabs.harvard.edu/abs/2015MNRAS.454.1038R} {454, 1038}

\bibitem[\protect\citeauthoryear{{Rupke}, {Kewley}  \& {Chien}}{{Rupke}
  et~al.}{2010a}]{rupke2010}
{Rupke} D.~S.~N.,  {Kewley} L.~J.,   {Chien} L.,  2010a, preprint, \href
  {http://adsabs.harvard.edu/abs/2010arXiv1009.0761R} {} (\mn@eprint {arXiv}
  {1009.0761})

\bibitem[\protect\citeauthoryear{{Rupke}, {Kewley}  \& {Barnes}}{{Rupke}
  et~al.}{2010b}]{rupke2010a}
{Rupke} D.~S.~N.,  {Kewley} L.~J.,   {Barnes} J.~E.,  2010b, \mn@doi [\apjl]
  {10.1088/2041-8205/710/2/L156}, \href
  {http://adsabs.harvard.edu/abs/2010ApJ...710L.156R} {710, L156}

\bibitem[\protect\citeauthoryear{{S{\'a}nchez-Menguiano}, {S{\'a}nchez},
  {P{\'e}rez}  \& {Garc{\'{\i}}a-Benito}}{{S{\'a}nchez-Menguiano}
  et~al.}{2016}]{sanchez-menguiano2016}
{S{\'a}nchez-Menguiano} L.,  {S{\'a}nchez} S.~F.,  {P{\'e}rez} I.,
  {Garc{\'{\i}}a-Benito} 2016, \mn@doi [\aap] {10.1051/0004-6361/201527450},
  \href {http://adsabs.harvard.edu/abs/2016A%26A...587A..70S} {587, A70}

\bibitem[\protect\citeauthoryear{{S{\'a}nchez} et~al.,}{{S{\'a}nchez}
  et~al.}{2014}]{sanchez2014}
{S{\'a}nchez} S.~F.,  et~al., 2014, \mn@doi [A\&A]
  {10.1051/0004-6361/201322343}, \href
  {http://adsabs.harvard.edu/abs/2014A%26A...563A..49S} {563, A49}

\bibitem[\protect\citeauthoryear{{Scannapieco}, {Tissera}, {White}  \&
  {Springel}}{{Scannapieco} et~al.}{2005}]{scan05}
{Scannapieco} C.,  {Tissera} P.~B.,  {White} S.~D.~M.,   {Springel} V.,  2005,
  \mn@doi [MNRAS] {10.1111/j.1365-2966.2005.09574.x}, \href
  {http://adsabs.harvard.edu/abs/2005MNRAS.364..552S} {364, 552}

\bibitem[\protect\citeauthoryear{{Scannapieco}, {Tissera}, {White}  \&
  {Springel}}{{Scannapieco} et~al.}{2006}]{scan06}
{Scannapieco} C.,  {Tissera} P.~B.,  {White} S.~D.~M.,   {Springel} V.,  2006,
  \mn@doi [MNRAS] {10.1111/j.1365-2966.2006.10785.x}, \href
  {http://adsabs.harvard.edu/abs/2006MNRAS.371.1125S} {371, 1125}

\bibitem[\protect\citeauthoryear{{Scannapieco}, {Tissera}, {White}  \&
  {Springel}}{{Scannapieco} et~al.}{2008}]{scan08}
{Scannapieco} C.,  {Tissera} P.~B.,  {White} S.~D.~M.,   {Springel} V.,  2008,
  \mn@doi [MNRAS] {10.1111/j.1365-2966.2008.13678.x}, \href
  {http://adsabs.harvard.edu/abs/2008MNRAS.389.1137S} {389, 1137}

\bibitem[\protect\citeauthoryear{{Scannapieco}, {Wadepuhl}, {Parry}, {Navarro},
  {Jenkins}, {Springel}  \& {Teyssier}}{{Scannapieco} et~al.}{2012}]{scan12}
{Scannapieco} C.,  {Wadepuhl} M.,  {Parry} O.~H.,  {Navarro} J.~F.,  {Jenkins}
  A.,  {Springel} V.,   {Teyssier} R.,  2012, \mn@doi [\mnras]
  {10.1111/j.1365-2966.2012.20993.x}, \href
  {http://adsabs.harvard.edu/abs/2012MNRAS.423.1726S} {423, 1726}

\bibitem[\protect\citeauthoryear{{Sersic}}{{Sersic}}{1968}]{sersic1968}
{Sersic} J.~L.,  1968, {Atlas de galaxias australes}

\bibitem[\protect\citeauthoryear{{Stott} et~al.,}{{Stott}
  et~al.}{2014}]{stott2014}
{Stott} J.~P.,  et~al., 2014, \mn@doi [MNRAS] {10.1093/mnras/stu1343}, \href
  {http://adsabs.harvard.edu/abs/2014MNRAS.443.2695S} {443, 2695}

\bibitem[\protect\citeauthoryear{{Tacchella}, {Dekel}, {Carollo}, {Ceverino},
  {DeGraf}, {Lapiner}, {Mandelker}  \& {Primack Joel}}{{Tacchella}
  et~al.}{2016}]{tacchella2016}
{Tacchella} S.,  {Dekel} A.,  {Carollo} C.~M.,  {Ceverino} D.,  {DeGraf} C.,
  {Lapiner} S.,  {Mandelker} N.,   {Primack Joel} R.,  2016, \mn@doi [\mnras]
  {10.1093/mnras/stw131}, \href
  {http://adsabs.harvard.edu/abs/2016MNRAS.457.2790T} {457, 2790}

\bibitem[\protect\citeauthoryear{{Tinsley} \& {Larson}}{{Tinsley} \&
  {Larson}}{1978}]{tinsley1978}
{Tinsley} B.~M.,  {Larson} R.~B.,  1978, \mn@doi [\apj] {10.1086/156056}, \href
  {http://adsabs.harvard.edu/abs/1978ApJ...221..554T} {221, 554}

\bibitem[\protect\citeauthoryear{{Tissera}}{{Tissera}}{2000}]{tissera2000}
{Tissera} P.~B.,  2000, \mn@doi [ApJ] {10.1086/308774}, \href
  {http://adsabs.harvard.edu/abs/2000ApJ...534..636T} {534, 636}

\bibitem[\protect\citeauthoryear{{Tissera}, {Beers}, {Carollo}  \&
  {Scannapieco}}{{Tissera} et~al.}{2014}]{tissera2014}
{Tissera} P.~B.,  {Beers} T.~C.,  {Carollo} D.,   {Scannapieco} C.,  2014,
  \mn@doi [\mnras] {10.1093/mnras/stu181}, \href
  {http://adsabs.harvard.edu/abs/2014MNRAS.439.3128T} {439, 3128}

\bibitem[\protect\citeauthoryear{{Tissera}, {Pedrosa}, {Sillero}  \&
  {Vilchez}}{{Tissera} et~al.}{2016a}]{tissera2016a}
{Tissera} P.~B.,  {Pedrosa} S.~E.,  {Sillero} E.,   {Vilchez} J.~M.,  2016a,
  \mn@doi [\mnras] {10.1093/mnras/stv2736}, \href
  {http://adsabs.harvard.edu/abs/2016MNRAS.456.2982T} {456, 2982}

\bibitem[\protect\citeauthoryear{{Tissera}, {Machado}, {Sanchez-Blazquez},
  {Pedrosa}, {S{\'a}nchez}, {Snaith}  \& {Vilchez}}{{Tissera}
  et~al.}{2016b}]{tissera2016b}
{Tissera} P.~B.,  {Machado} R.~E.~G.,  {Sanchez-Blazquez} P.,  {Pedrosa} S.~E.,
   {S{\'a}nchez} S.~F.,  {Snaith} O.,   {Vilchez} J.,  2016b, \mn@doi [\aap]
  {10.1051/0004-6361/201628188}, \href
  {http://adsabs.harvard.edu/abs/2016A%26A...592A..93T} {592, A93}

\bibitem[\protect\citeauthoryear{{Torrey}, {Vogelsberger}, {Sijacki},
  {Springel}  \& {Hernquist}}{{Torrey} et~al.}{2012}]{torrey2012}
{Torrey} P.,  {Vogelsberger} M.,  {Sijacki} D.,  {Springel} V.,   {Hernquist}
  L.,  2012, \mn@doi [\mnras] {10.1111/j.1365-2966.2012.22082.x}, \href
  {http://adsabs.harvard.edu/abs/2012MNRAS.427.2224T} {427, 2224}

\bibitem[\protect\citeauthoryear{{Tremonti} et~al.,}{{Tremonti}
  et~al.}{2004}]{tremonti2004}
{Tremonti} C.~A.,  et~al., 2004, \mn@doi [ApJ] {10.1086/423264}, \href
  {http://adsabs.harvard.edu/abs/2004ApJ...613..898T} {613, 898}

\bibitem[\protect\citeauthoryear{{Woods}, {Geller}  \& {Barton}}{{Woods}
  et~al.}{2006}]{woods2006}
{Woods} D.~F.,  {Geller} M.~J.,   {Barton} E.~J.,  2006, \mn@doi [\aj]
  {10.1086/504834}, \href {http://adsabs.harvard.edu/abs/2006AJ....132..197W}
  {132, 197}

\bibitem[\protect\citeauthoryear{{Woosley} \& {Weaver}}{{Woosley} \&
  {Weaver}}{1995}]{WW95}
{Woosley} S.~E.,  {Weaver} T.~A.,  1995, \mn@doi [ApJS] {10.1086/192237}, \href
  {http://adsabs.harvard.edu/abs/1995ApJS..101..181W} {101, 181}

\bibitem[\protect\citeauthoryear{{Wuyts}, {Wisnioski}, {Fossati}, {F{\"o}rster
  Schreiber}  \& ...}{{Wuyts} et~al.}{2016}]{wuyts2016}
{Wuyts} E.,  {Wisnioski} E.,  {Fossati} M.,  {F{\"o}rster Schreiber} N.~M.,
  ... 2016, \mn@doi [\apj] {10.3847/0004-637X/827/1/74}, \href
  {http://adsabs.harvard.edu/abs/2016ApJ...827...74W} {827, 74}

\bibitem[\protect\citeauthoryear{{Yuan}, {Kewley}, {Swinbank}, {Richard}  \&
  {Livermore}}{{Yuan} et~al.}{2011}]{yuan2011}
{Yuan} T.-T.,  {Kewley} L.~J.,  {Swinbank} A.~M.,  {Richard} J.,   {Livermore}
  R.~C.,  2011, \mn@doi [ApJL] {10.1088/2041-8205/732/1/L14}, \href
  {http://adsabs.harvard.edu/abs/2011ApJ...732L..14Y} {732, L14}

\bibitem[\protect\citeauthoryear{{Zaritsky}, {Kennicutt}  \&
  {Huchra}}{{Zaritsky} et~al.}{1994}]{zaritsky1994}
{Zaritsky} D.,  {Kennicutt} Jr. R.~C.,   {Huchra} J.~P.,  1994, \mn@doi [ApJ]
  {10.1086/173544}, \href {http://adsabs.harvard.edu/abs/1994ApJ...420...87Z}
  {420, 87}

\bibitem[\protect\citeauthoryear{{van Zee}, {Salzer}, {Haynes}, {O'Donoghue}
  \& {Balonek}}{{van Zee} et~al.}{1998}]{van-zee1998}
{van Zee} L.,  {Salzer} J.~J.,  {Haynes} M.~P.,  {O'Donoghue} A.~A.,
  {Balonek} T.~J.,  1998, \mn@doi [\aj] {10.1086/300647}, \href
  {http://adsabs.harvard.edu/abs/1998AJ....116.2805V} {116, 2805}

\makeatother
\end{thebibliography}
\bibliographystyle{mnras}

% Definiciones solo para el MNRAS para ApJ no ponerlas
\def\apj{ApJ}
\def\apjl{ApJ}
\def\aj{AJ}
\def\mnras{MNRAS}
\def\aa{A\&A}
\def\nat{Nature}
\def\araa{ARA\&A}
\def\aap{A\&A}

\IfFileExists{\jobname.bbl}{}
{
\typeout{}
\typeout{****************************************************}
\typeout{****************************************************}
\typeout{** Please run "bibtex \jobname" to optain}
\typeout{** the bibliography and then re-run LaTeX}
\typeout{** twice to fix the references!}
\typeout{****************************************************}
\typeout{****************************************************}
\typeout{}
}

% Don't change these lines
\bsp  % typesetting comment
\label{lastpage}
\end{document}